\documentclass[prl,reprint,superscriptaddress]{revtex4-1}

\usepackage{amsmath,amssymb}
\usepackage{verbatim}
\usepackage{graphicx}
\usepackage{hyperref}
\usepackage{color}
\usepackage{mathrsfs}
\usepackage{slashed}

\usepackage[utf8]{inputenc}

\newcommand {\cL}{{\cal L}}

\newcommand {\cN}{{\cal N}}

\def\a{\alpha}
\def\b{\beta}

\def\d{\delta}
\def\e{\epsilon}
\def\f{\phi}
\def\g{\gamma}

\def\l{\lambda}

\def\p{\pi}

\def\x{\xi}

\newcommand{\be}{\begin{equation}}
	\newcommand{\ee}{\end{equation}}
\newcommand{\bea}{\begin{eqnarray}}
	\newcommand{\eea}{\end{eqnarray}}

\newcommand{\ba}{\begin{array}}
	\newcommand{\ea}{\end{array}}

\def\double #1{#1{\hbox{\kern-2pt $#1$}}}

\newcommand{\bsubeq}{\begin{subequations}}
	\newcommand{\esubeq}{\end{subequations}}

\def\ft#1#2{{\textstyle{\tfrac{\scriptstyle #1}{\scriptstyle #2} } }}

\def\rmi{{\rm i}}

\def\a{\alpha}
\def\b{\beta}
\def\g{\gamma}

\def\d{\delta}

\def\e{\epsilon}

\def\f{\phi}

\def\p{\psi}

\def\l{\lambda}

\def\x{\xi}

\begin{document}
	
	\title{Supersymmetric Carroll Galileons in Three Dimensions}

 \author{Utku Zorba}
	\email{utku.zorba@iuc.edu.tr}
	\affiliation{Department Of Engineering Sciences, Istanbul University-Cerrahpasa,\\
		  Avcilar 34320 Istanbul, T\"urkiye}
	
		\author{Ilayda Bulunur}
	\email{bulunur@itu.edu.tr}
	\affiliation{Department of Physics,
		Istanbul Technical University,
		Maslak 34469 Istanbul,
		T\"urkiye}

	\author{Oguzhan Kasikci}
	\email{kasikcio@itu.edu.tr}
	\affiliation{Department of Physics,
		Istanbul Technical University,
		Maslak 34469 Istanbul,
		T\"urkiye}
	
	\author{Mehmet Ozkan}
	\email{ozkanmehm@itu.edu.tr}
	\affiliation{Department of Physics,
		Istanbul Technical University,
		Maslak 34469 Istanbul,
		T\"urkiye}
	
	\author{Yi Pang}
	\email{pangyi1@tju.edu.cn}
	\affiliation{Center for Joint Quantum Studies and Department of Physics,\\
		School of Science, Tianjin University, Tianjin 300350, China }
	\affiliation{Peng Huanwu Center for Fundamental Theory,\\
		Hefei, Anhui 230026, China }
	
		\author{Mustafa Salih Zog}
	\email{zog@itu.edu.tr}
	\affiliation{Department of Physics,
		Istanbul Technical University,
		Maslak 34469 Istanbul,
		T\"urkiye}

	\date{\today}
	

	\begin{abstract}
	
We present the first example of an interacting Carroll supersymmetric field theory with both temporal and spatial derivatives, belonging to the Galileon class, where the non-linear field equation remains second-order in derivative. To achieve this, we introduce two novel tools. First, we generalize the bosonic map between the
Galilei and Carroll algebras to include supersymmetry by using a spinor basis for the symmetry generators. We then show that Carroll superalgebras are naturally connected to Euclidean, rather than Poincar\'e, superalgebras. Using the real multiplet of the three-dimensional $\cN=1$ Euclidean supersymmetry, we construct the scalar multiplet for $\cN=2$ Carroll supersymmetry and develop a tensor calculus to realize the aforementioned model. These results offer new insights into the structure of genuine higher-dimensional Carroll field theories and Carroll supersymmetry. While these tools are utilized to build a specific model, we anticipate that they possess broader applications in Carrollian physics.
		
	\end{abstract}
	
	
	\maketitle
	\allowdisplaybreaks
	
	 \textit{Introduction} --  Galilei and Carroll symmetries represent two distinct limits of the relativistic symmetry, with the former arising when the speed of light $c$ tends to infinity and the latter when $c$ approaches zero \cite{Leblond1965,Gupta1966}. While these symmetries emerge as limits of relativistic models, they possess intrinsic interest and significance. On one hand, Galilei symmetry and its extensions have been extensively utilized in formulating effective theories of fractional quantum Hall effect \cite{Son:2013rqa,Geracie:2014nka,Gromov:2014vla}, as well as in Lifshitz and Schr{\"o}dinger holography \cite{Son:2008ye,Balasubramanian:2008dm,Christensen:2013lma, Christensen:2013rfa, Hartong:2014oma}. On the other hand, Carroll symmetry is particularly relevant to the physics of null hypersurfaces \cite{Duval:2014uoa, Duval:2014uva, Duval:2014lpa, Ciambelli:2018wre, Figueroa-OFarrill:2021sxz, Herfray:2021qmp, Mittal:2022ywl, Campoleoni:2023fug, Penna:2015gza}, particularly to the black hole horizon \cite{Penna:2018gfx, Donnay_2019, Ciambelli:2019lap, Redondo_Yuste_2023,Freidel:2022vjq,Gray:2022svz, Bicak:2023rsz}, condensed matter systems and fractons \cite{Bidussi:2021nmp, Grosvenor:2021hkn, Figueroa_O_Farrill_2023, Figueroa-OFarrill:2023qty, Bagchi:2022eui, Marsot:2022imf}, hydrodynamics \cite{deBoer:2017ing, Ciambelli:2018xat,Campoleoni:2018ltl,Petkou:2022bmz,Freidel:2022bai,Bagchi:2023ysc}, cosmology \cite{deBoer:2021jej, Avila:2023brd} and celestial and conformal Carroll holography \cite{Raclariu:2021zjz, Pasterski:2021rjz, Chen:2021xkw, Donnay:2022aba, Bagchi:2022emh, Donnay:2022wvx, Chen:2023naw}. 
	 
	 A distinctive feature of these symmetries is that the spatial derivative is inert under Galilei boosts, while the time derivative is inert under Carroll boosts \cite{Leblond1965}. The way that the partial derivatives transform has significant implications on the structure of field theories invariant under the two symmetries. For instance, at the two-derivative level, Galilei-invariant theories cannot contain time derivatives -- indeed the action for the free Schr{\"o}dinger equation is in fact invariant under the Schr{\"o}dinger group which contains the mass, dilatation and the expansion apart from the Galilei group -- while Carroll-invariant models lack spatial derivatives, unless one uses a reducible representation of the relevant symmetry groups \cite{deBoer:2023fnj}. In this sense, Carroll-invariant models can be viewed as an infinite collection of identical one-dimensional models, each labeled by the spatial coordinates \cite{deBoer:2023fnj, Banerjee:2023jpi, Kasikci:2023zdn}. These observations suggest that Galilei and Carroll field theories exhibit dynamics effectively ``frozen” in time or space. In this Letter, we shall focus on Carrollian invariant models for which the aforementioned frozen dynamics is merely an illusion due to our limited knowledge based on the two-derivative Lagrangian.  

 A class of single-field Carrollian theories with both types of derivatives is the $c\to 0$ limit of $D$-dimensional relativistic Galileons of degree $N$
 \begin{align}
\cL_{CGal}^N & = \lim_{c \to 0} \left( c^2 \f \, \d^{A_1 \ldots A_N}_{B_1 \ldots B_N} \prod_{a=1}^N S_{A_a}{}^{B_a} \right)\,, 
\label{CarrollGalileons}
\end{align}	
 where $A_1,B_2, \ldots, A_N, B_N=0,1,\ldots,D-1$ and we define $S_{AB} = \partial_A \partial_B \f$. Furthermore, we introduce an overall factor of $c^2$ to avoid possible rescaling of the scalar field $\f$. We refer to these models as Carroll Galileons. These models bypass higher time derivatives, and therefore the Ostrogradsky instability, see \cite{Tadros:2024fgi} for a recent study, and, as we will discuss shortly, these theories also exhibit a rich vacuum structure.

 Inclusion of supersymmetry is a natural tool to generalize this notion to fermions, and is the first step towards the construction of such superconformal field theories needed for exploring the flat-space holography via superBMS / superconformal Carroll field theories \cite{Bagchi:2022owq}. Furthermore, supersymmetry could provide deeper insights into this intriguing behavior and its implications by improving their quantum behavior and offering tools to solve these models. However,  $\mathcal{N} = 1$ supersymmetry would completely banish the spatial derivative terms, and the remaining action would contain only time derivatives, reverting back to the standard form of Carrollian theories since supercharges square to time translation \cite{Kasikci:2023zdn}. This fact has recently been shown to lead to a Carrollian supersymmetric higher-derivative theory without a purely bosonic sector \cite{Kasikci:2023zdn}, in contrast to Galilean supersymmetric models, which typically retain a purely bosonic sector. Inspired by the centrally extended Galilei superalgebra, known as the Bargmann superalgebra \cite{Gomis:2004pw}, one potential avenue for introducing spatial derivatives in Carrollian supertransformations is through the use of extended supersymmetry. However, constructing Carrollian extended superalgebras and their multiplets is a highly nontrivial task. A naive $c \to 0$ limit removes again the spatial translations in the square of the supercharges \cite{Bergshoeff:2015wma}. As far as we are aware, there is no established method of deriving such algebras from relativistic algebras or non-relativistic counterparts. 
 
	 Here, we address this challenge by introducing two new tools with potentially broader implications beyond this paper. First, we generalize the bosonic map between the Galilei and Carroll algebras \cite{Bergshoeff:2022qkx} to include supersymmetry by using a spinor basis for the symmetry generators. This provides a new tool for constructing $\cN > 1$ Carrollian superalgebras. Second, we show that the ultra-relativistic superalgebra derived from this relation also arises as the $c \to 0$ limit of the Euclidean superalgebra. This contrasts with the nature of non-relativistic supersymmetry, which emerges as the $c \to \infty$ limit of the Poincaré superalgebra. We finally construct the scalar superfield for the three-dimensional $\mathcal{N}=2$ Carroll superalgebra and obtain the supersymmetric completion of the simplest interacting Carroll Galileon model, corresponding to $N=2$ in \eqref{CarrollGalileons}. This represents  the first Carrollian supersymmetric model with spatial derivatives.

	 \textit{Non-/Ultra-Relativistic Duality in Spinor Basis} --  In two spacetime dimensions, it is possible to start with the generators of the Galilei (or Carroll) algebra and produce the generators of the Carroll (or Galilei) algebra by means of a coordinate transformation $t \leftrightarrow x$ \cite{Duval:2014uoa} (see \cite{Ravera:2022buz} for its supersymmetric extension). To see that, consider the generators of the Galilei algebra 
	 \begin{align}
	 	H & = \partial_t \,, & P & = \partial_x \,, & G & = t \partial_x \,.
	 \end{align}
	It is evident that $H$ and $P$ exchange roles as $t \leftrightarrow x$ and the Galilei boost $G$ becomes the Carroll boost $C = x \partial_t$. To extend this duality beyond two dimensions, let us first consider the spacetime split Poincar\'e algebra, where the $D-$dimensional generators, coordinates $x^A$ and the gamma matrices $\gamma^A$ are split into temporal and spatial parts, labeled by $0$ and $i = 1, \dots, D-1$, respectively. Omitting the spatial rotations, in a spinor basis with the following definitions
	\begin{align}
	\widehat H & = \g^0 H \,, & \widehat P & = \g^i P_i \,, & \widehat J &=  \g^{i0} J_{i0} \,,
	\end{align}
where $H \equiv P_0$ generates time translation, $P_i$ denotes the spatial translation and $J_{i0}$ corresponds to the Lorentz boost. Accordingly, we may choose the differential representation  
\begin{align}
	\widehat H & = \g^0 \partial_0 \,, & \widehat P & = \g^i \partial_i \,, & \widehat J& = \g^i \g^0 x_i \partial_0 + \g^0 \g^i  x_0 \partial_i \,.
\end{align}
Let us now consider the following map
\begin{align}
 \gamma^{i} x_{i} & \leftrightarrow \g^0 x_0 \,, &  \gamma^{i} \partial _{i} & \leftrightarrow \gamma^0 \partial_0  \,. 
\label{NewTransformation}
\end{align}
As in $D=2$, this map indicates interchanging roles of $\widehat H$ and $\widehat P$. Although $\widehat J$ remains invariant, a closer look reveals that the invariance is achieved as the Galilei part $ \widehat{G} = \g^0 \g^i x_i \partial_0$  turns into the Carroll part $\widehat C = \g^i \g^0 x_0 \partial_i$ and vice versa. We note that the invariance under \eqref{NewTransformation} is also present in the supersymmetry generators, because implementing \eqref{NewTransformation} in
\begin{equation}
	Q_\a = \tfrac{\partial}{\partial \bar\theta^\a}  -  \left(\gamma^0 \theta \right)_\alpha \partial_0  -  \left(\gamma^i \theta \right)_\alpha \partial_i \,, 
\end{equation}
transforms its Galilei part  ($\gamma^i \partial_i$) and Carroll part  ($\gamma^0 \partial_0$) into each other. 

With this property of super-Poincar\'e algebra in mind, we now turn to the generators of three-dimensional $\cN = 2$ Galilei algebra, which contains the usual Galilei generators as well as two SUSY generators  $(Q^G_+,\, Q^G_-)$. In particular, the SUSY generators are given as \cite{Gomis:2004pw}
\begin{align}
Q^G_+ & = \tfrac{\partial}{\partial \bar \theta_+}  - \tfrac12 \g^i \theta_- \partial_{i } + \tfrac12 \g_0   \theta_+ \partial_{t}\,, \nonumber\\
Q^G_- & = \tfrac{\partial}{\partial \bar \theta_-}  -  \tfrac12  \g^i \theta_+ \partial_{i} \,,
\label{NRN2SUSY}
\end{align}
where we suppress the spinor indices. These generators are two-component Majorana spinors, and they obey the following anti-commutation relations 
\be
\{Q_+^G, Q_+^G\} = - \g_0 H,\quad \{Q_+^G, Q_-^G\} = \g^i P_i\, .
\ee
Now we can apply the map \eqref{NewTransformation} to obtain a representation for the three-dimensional $\cN=2$ Carroll supersymmetry generators. In particular, the supersymmetry generators are given by
\begin{align}
Q_+ & = \tfrac{\partial}{\partial \bar \theta_+}  -  \tfrac12 \g^i   \theta_+ \partial_{i} + \tfrac12 \g_0 \theta_- \partial_{t} \,, \nonumber\\
Q_- & = \tfrac{\partial}{\partial \bar \theta_-}  + \tfrac12 \g_0 \theta_+ \partial_{t} \,.
\label{N2CarrollSUSYGenerators}
\end{align}
Together with the generators of bosonic Carroll algebra, they span the three-dimensional $\cN=2$ Carroll superalgebra of the form
\begin{align}
\left[ J,P_i \right] & =  \e_{ij} P^j \,, & \left[ J,C_i \right] & =  \e_{ij} C^j \,, \nonumber\\
 \left[ P_i,C_j \right] & = \d_{ij} H \,, & \left[ C_i,Q_+ \right] & = \tfrac{1}{2} \g_i Q_-\,,\nonumber\\
  \left[ J,Q^\pm \right] & = \tfrac{1}{2} \g_0 Q^\pm \,,  &   \{Q_+,Q_+ \} & = \g^i P_i \,, \nonumber\\
  \{Q_+ ,Q_- \} &=   - \g_0 H \,,
\label{N2Carroll}
\end{align}
where we used the two-dimensional identity for the spatial rotation $J _{ij}= \e_{ij} J$. Note that in the absence of $J$, we have the Carroll-boost superalgebra, which is the dual of the Galilei-boost superalgebra in spinor basis.  As desired, this algebra contains spatial translations, which is crucial to the construction of Carroll supersymmetric models with spatial derivatives. Although here we utilize the map \eqref{NewTransformation} to obtain the three-dimensional $\cN=2$ Carroll supersymmetry, as mentioned previously, the map \eqref{NewTransformation} is quite generic, and enables one to construct ultra-relativistic theories from non-relativistic ones. 

 \textit{$\cN = 2$ Carroll Superalgebra from $\cN=1$ Euclidean Superalgebra} --  The structure of supersymmetry algebra in \eqref{N2Carroll} provides an important insight into the origin of Carroll superalgebra. While both Galilei and Carroll groups can emerge as non-relativistic and ultra-relativistic limits, respectively, of the Poincar\'e group, the inclusion of fermionic generators brings in a crucial difference. The Carroll superalgebra prefers a Euclidean signature, rather than a Lorentzian one. For the purely bosonic  commutation relations, this choice would not change the structure constants. For the fermionic sector, consider the anticommutator of $\cN=1$ Euclidean supersymmetry generator
 \begin{equation}
\{Q, Q^\dagger \} = \tfrac{2}{c} \g^E_0 H + 2 \g^i P_i \,.
 \end{equation}
where $Q$ is a two-component Dirac spinor and $(\g^E_0)^2=1$. Decomposing the Dirac spinor $Q$ into real spinors $Q_{\pm}$ via $Q = Q_+ + \rmi Q_-$ gives rise to
\begin{align}
\{Q_\pm, Q_\pm\} &=   \g^i P_i \,, & \{Q_+, Q_-\}  &= \tfrac{\rmi}{c} \g^E_0 H \,,
\label{PreCarroll}
\end{align}
 which precisely recover the Carroll superalgebra in the $c \to 0$ limit once $Q_-$ is rescaled as $Q_- \to Q_-/c$. Note that to match the convention of \eqref{N2Carroll}, we need to redefine $\g^E_0 \to \rmi \g_0$ since in \eqref{N2Carroll} $\g_0^2 = - 1$ due to its relativistic origin. One may wonder what if we try to obtain the algebra \eqref{PreCarroll} from the $\cN=2$ super-Poincar\'e algebra. For instance, one can decompose the relativistic supersymmetry generator as $Q = Q_1 + \rmi Q_2$, followed by defining $Q_\pm$ as $Q_\pm =(Q_1 \pm \rmi \g_0 Q_2) /2 $, giving rise to the same structure of \eqref{PreCarroll}. Note that in this case,  $Q_\pm$ are Dirac spinors due to the appearance of the factor $\rmi$. This is in contrast to the Euclidean case where the $Q_\pm$ are real spinors. Nonetheless, this is harmless before the ultra-relativistic limit is taken, since $Q_\pm$ are complex conjugate to each other. However, once the limit is taken, they are no longer conjugate pair and one ends up with two independent Dirac spinors as supersymmetry generators. This appears to be different from the $\cN=2$ structure given in \eqref{N2Carroll}, which is realized with Majorana spinors.

The relation between the $\cN=1$ Euclidean and the $\cN=2$ Carroll supersymmetry suggests that we may consider the following real superfield, which is the reformulation of the real superfield of three-dimensional $\cN = 1$ Euclidean supersymmetry \cite{McKeon:2001su}
\begin{align}
	\Phi & =    \phi + \bar\theta_+ \l_1 + \bar\theta_- \l_2 + \tfrac12 \bar\theta_+ \theta_+ F_1 + \bar\theta_+ \theta_- F_2 \nonumber\\
	&  + \tfrac12 \bar\theta_- \theta_- F_3 + \bar\theta_+ \g_0  \theta_- F_4 + \bar\theta_+ \g^i \theta_- B_i  \nonumber\\
	&  + \tfrac12 \bar\theta_+ \theta_+ \bar\theta_- \p_1  + \tfrac12 \bar\theta_- \theta_- \bar\theta_+ \p_2 + \tfrac14 \bar\theta_+ \theta_+ \bar\theta_- \theta_- S \,.
\end{align}
This is also a superfield of $\cN=2$ Carroll superalgebra since its structure is not deformed due to rescaling the fields and $\theta_\pm$ \cite{Kasikci:2023zdn}. Thus, using the operator representation of the supersymmetry generators \eqref{N2CarrollSUSYGenerators}, the transformation rules for this multiplet can be found by using the standard rule
\begin{eqnarray}
    \d \Phi =  [\bar\e_+ Q_+, \Phi] +  [\bar\e_- Q_-, \Phi] \,.
\end{eqnarray}
The resulting transformation rules are provided in the Supplemental Material, and, as promised, these transformation rules do involve spatial derivatives that is essential for the construction of Carroll supersymmetric models that cannot be treated as a one-dimensional model. This long multiplet can be shortened by imposing the condition $D_- \Phi = 0$ where
\begin{equation}
	D_-  = \rmi  \Big( \tfrac{\partial}{\partial \bar\theta_{-}} - \tfrac{1}{2}  \g_0 \theta_+ \partial_{t} \Big) ,\quad \{D_-,\,Q_{\pm}\}=0\ ,
\end{equation}
which corresponds to the following consistent truncation
\begin{align}
\l_2 & = 0 \,, & F_4 & = -\tfrac12\dot\f \,, & \p_1 & = - \tfrac12 \g_0 \dot\l_1 \,, & S & = -\tfrac14 \ddot \f \,,\nonumber\\
\p_2 & =  0\,, & F_2 & = 0\,, &F_3 & = 0 \,, & B_i & = 0 \,.
\end{align}
The components of this scalar multiplet is therefore, given by $(\f, \l, F)$ with the transformation rules
\begin{eqnarray}
	\d \f & = &  \bar\e_+ \l \,,\nonumber\\
	\d \l & =& \tfrac12 \g^i \partial_i \f \e_+ - \g_0 \dot \f \e_- + F \e_+  \,,\nonumber \\
	\d F &=& \tfrac12 \bar\e_+ \g^i \partial_i \l  - \bar\e_- \g_0 \dot \l \,.
	\label{N2ScalarMultiplet}
\end{eqnarray}
It is important to note that under the map \eqref{NewTransformation}, the scalar multiplet \eqref{N2ScalarMultiplet} becomes identical to the scalar multiplet of the $\cN = 2$ Galilei superalgebra \cite{Bergshoeff:2015ija}. The scalar multiplet obeys the following composition rule
\begin{eqnarray}
    \phi_3 &=& \phi_1 \phi_2 \,,\nonumber\\
    \l_3 &=& \phi_1 \lambda_2 + \phi_2 \lambda_1  \,,\nonumber\\
    F_3 &=& \phi_ 1 F_2 + \phi_2 F_1 - \bar\l_1 \l_2 \,,
    \label{ScalarMultiplet}
\end{eqnarray}
where for each $I=1,2,3$, $(\phi_I, \l_I, F_I)$  forms a scalar multiplet. Since the highest component of a scalar multiplet transforms to total derivative, it can be utilized as a consistent action principle. Hence, we consider
 \begin{eqnarray}
     \cL &=& \phi F^\prime + \phi^\prime F - \bar\l \l^\prime \,.
 \end{eqnarray}
as an invariant action. Based on this action, we may construct the simplest dynamical model and the supersymmetric completion of the $N=2$ model in \eqref{CarrollGalileons} given by
  \begin{equation}
\cL =- \tfrac13 \phi \big( \ddot\f\,\partial^i \partial_i \f - \partial_i \dot \f \partial^i \dot \f \big) \,,
      \label{STFracton}
  \end{equation}
 First, choosing the primed scalar multiplet as $(\ddot \phi, \ddot \l, \ddot F)$, we arrive at the simplest dynamical Lagrangian
\begin{eqnarray}
    \cL &=& 2  F \ddot \phi - \bar\l \ddot \l \,.
    \label{SimplestDynamical}
\end{eqnarray}
This does not contain the $\dot\phi^2$ term, which is not allowed by the transformation rules \eqref{ScalarMultiplet}. To construct the supersymmetric completion of \eqref{STFracton}, we first map the highest component of one scalar multiplet to the lowest component of a second one, which is a standard procedure in tensor calculus to form the kinetic multiplet. We then apply the composition rule and obtain the supersymmetric completion of the bosonic model \eqref{STFracton}
\begin{equation}
\cL = -2 F^2 \ddot \phi +\tfrac12 \dot\f^2 \partial_i \partial^i \f -2 F \dot{\bar \l} \dot \l -2 \dot \f \dot{\bar\l} \g^i \partial_i \l \,,
\label{SuperSwifton}
\end{equation}
where we performed various partial integration for simplification, and rescaled the Lagrangian with an overall factor of $-4/3$. Upon setting $F$ and $\lambda$ to $0$ using their field equations, the model reduces to the purely bosonic one given in \eqref{STFracton}. We note that both \eqref{SimplestDynamical} and \eqref{SuperSwifton} exhibit a possible fermionic Galileon structure with a shift symmetry $\l \to \l + \x$ where $\x$ is a constant spinor. Furthermore, their field equations do not exhibit a higher time-derivative structure. The higher-derivative model \eqref{SuperSwifton} is also the first example of a Carroll supersymmetric model that includes spatial derivatives. For the reader's convenience, we give the details of the calculations in the Supplemental Material. To study vacuum solution of \eqref{SuperSwifton}, we first set $F=0$ using its field equation. The equation of motion of $\phi$ is then given by 
\be
\partial_i\dot\phi\partial^i\dot\phi-\ddot\phi\partial_i\partial^i\phi=0\ ,
\ee
which admits special solutions of the form $\phi=\a t+\b(x,y)$ where $\a$ is a constant and $\b(x,y)$ is an arbitrary function of spatial coordinates. Considering perturbations around this particular vacuum, $\phi=\a t+\b(x,y)+\delta\phi$, the quadratic action for the perturbation is of the form 
\be
\delta{\cal L}^{(2)}=\ft12\partial_i\partial^i\b\delta\dot\phi^2+\a\delta\dot\phi\partial_i\partial^i\delta\phi\ .
\ee
One can easily compute the Hamiltonian and see that it is positive definite when $\partial_i\partial^i\b>0$. The equation above should allow for more vacuum solutions which we leave for future investigations. 
	
	\textit{Conclusions and outlook.}--- In this Letter, we have constructed the first supersymmetric Carroll invariant model with spatial derivatives, belonging to a class of models that we refer to as the Carroll Galileons. To achieve this, we developed two novel techniques that are expected to have broader implications beyond the scope of this work. First, we extend the well-known two-dimensional duality between  Carroll and Galilei boost algebras to arbitrary dimensions by introducing a spinor-based framework. Utilizing this tool, we uncover differential representations and the commutation relations for three-dimensional $\cN=2$ Carroll superalgebra. We then demonstrated that Carroll superalgebras are related to Euclidean superalgebras rather than Poincar\'e superalgebra. Using the real multiplet of three-dimensional $\cN=1$ Euclidean supersymmetry, we constructed the corresponding $\cN=2$ Carroll real multiplet, and a scalar multiplet. Finally, using tensor calculus, we established the first supersymmetric Carrollian model with spatial derivatives, representing the supersymmetric completion of the Galileon model given in \eqref{STFracton}. 

An exciting avenue for future work would be to construct the conformal extension of the $\cN=2$ Carroll superalgebra and find its relation to a super-BMS algebra. To our knowledge, such relation only exists for conformal Carroll algebras when supercharges square to the Hamiltonian \cite{Bagchi:2022owq}. Once this is achieved, it would indeed be desirable to understand how this super-BMS arises as an asymptotic symmetry algebra at null infinity, following the analysis of \cite{Lodato:2016alv}.

Finally, it would be interesting to extend our work to build $\cN=2$ Carroll supergravities and study supersymmetric black holes in these theories. We recall that in the relativistic case, supersymmetric localization technique has been utilized to understand the microstates of supersymmetric black holes \cite{Benini:2015eyy} via holography. It is desirable to extend these works to supersymmetric Carrollian black holes and provide a precise counting of their entropy resolving the puzzle arising from naively taking the $c\rightarrow 0$ of the relativistic result \cite{Ecker:2023uwm}. 
	
	\textit{Acknowledgements.}---  M.O. is supported in part by TUBITAK grant 121F064, the Distinguished Young Scientist Award BAGEP of the Science Academy, and the Outstanding Young Scientist Award of the Turkish Academy of Sciences (TUBA-GEBIP). The work of Y.P. is supported by the National Natural Science Foundation of China (NSFC) under grant No. 12175164 and the National Key Research and Development Program under grant No. 2022YFE0134300. This work is also partially supported by Peng Huanwu Center for Fundamental Theory, under grant No.12247103.

	\bibliographystyle{utphys}
	\bibliography{ref}

 \clearpage

\appendix{}

\begin{center}\textbf{SUPPLEMENTAL MATERIAL\\
(APPENDICES)}\end{center}

In the Supplemental Material to our letter, we present the full supersymmetry transformation rules for the three-dimensional $\cN=2$ real multiplet, using the operator representations of $Q_\pm$. We also provide the tensor calculus that is utilized for the construction of an action principle for $\cN=2$ Carroll supersymmetric interacting scalar field theory that involves both the spatial and the time derivatives. 
 
\section{Transformation Rules for the Real Multiplet}

The supersymmetry transformation rules for the real multiplet are given by
\begin{align}
	\delta \phi= & \bar{\epsilon}_{+} \lambda_1 +\bar{\epsilon}_-\lambda_2\ ,
 \nonumber\\
	\delta \lambda_1= & \tfrac12 \gamma^i \partial_i \phi \epsilon_{+}+F_1 \epsilon_{+}-\tfrac12 \g_0\dot{\phi}\e_-+F_2\e_- +\g_0F_4\e_-  \nonumber\\
	& +\g^iB_i\e_-\ , 
 \nonumber\\
	\delta \lambda_2= & -\tfrac12 \gamma_0 \dot{\phi} \epsilon_+ +F_2 \epsilon_{+}-\gamma_0 F_4 \epsilon_{+}-\gamma^i B_i \epsilon_{+} + F_3 \epsilon_{-}\ , 
 \nonumber\\
	\delta F_1= &\tfrac12 \bar{\epsilon}_{+} \gamma^i \partial_i \lambda_1+\bar{\e}_-\p_1-\tfrac12\bar{\e}_- \g_0\dot{\l}_1\ ,
 \nonumber\\
	\delta F_2= & -\tfrac14 \bar{\epsilon}_+ \gamma_0 \dot{\lambda}_1+\tfrac14 \bar{\epsilon}_{+} \gamma^i \partial_i \lambda_2-\tfrac12 \bar{\epsilon}_{+} \psi_1 - \dfrac{1}{4}\epsilon_- \gamma_{0} \dot{\lambda_2} \nonumber\\
	&  - \dfrac{1}{2} \epsilon_- \psi_2\ , 
   \nonumber\\
	\delta F_3= & -\tfrac12 \bar{\epsilon}_{+} \gamma_0 \dot{\lambda}_2+\bar{\epsilon}_{+} \psi_2\ ,\nonumber\\
	\delta F_4= & -\tfrac14 \bar{\epsilon}_+ \dot{\lambda}_1+\tfrac14 \bar{\epsilon}_{+} \gamma^{0 i} \partial_i \lambda_2-\tfrac12 \bar{\epsilon}_{+} \gamma_0 \psi_1 + \dfrac{1}{4} \bar{\epsilon}_-\dot\lambda_2 \nonumber\\
	&+ \dfrac{1}{2}\bar{\e}_-\gamma_{0}\psi_2\ , \nonumber\\
	\delta B_i= & \tfrac14 \bar{\epsilon}_+ \gamma_{i 0} \dot{\lambda}_1-\tfrac14\bar{\epsilon}_{+} \partial_i \lambda_2+\tfrac14\bar{\epsilon}_{+}\gamma_i^{\ j} \partial_j \lambda_2+\tfrac12 \bar{\epsilon}_{+} \gamma_i \psi_1  \nonumber\\
	&- \dfrac{1}{4} \bar{\e}_-\gamma_{i 0}\dot{\lambda}_2 - \dfrac{1}{2} \bar{\e}_-\gamma_{i}\psi_2\ , \nonumber\\
	\delta \psi_1= & -\tfrac12 \gamma_0 \dot{F}_1 \epsilon_{+}-\tfrac12 \gamma^i \partial_i F_2 \epsilon_{+}+\tfrac12 \gamma^{i 0} \partial_i F_4 \epsilon_{+}  \nonumber\\
	& +\tfrac12\partial^i B_i\epsilon_{+}  -\tfrac12\gamma^{ij} \partial_i B_j\epsilon_{+} + \tfrac12 \gamma_{0}\dot{F_2}\e_- + \tfrac12 \dot{F}_4\e_-  \nonumber\\
	& + \tfrac12 \gamma^{i 0}\dot{B}_{i}\e_- +  S\e_-\ ,
  \nonumber\\
	\delta \psi_2= & \tfrac12 \gamma_0 \dot{F}_2 \epsilon_{+}+\tfrac12 \gamma^i \partial_i F_3 \epsilon_{+}-\tfrac12 \dot{F}_4 \epsilon_{+}+\tfrac12 \gamma^{0i} \dot{B}_i \epsilon_{+}  \nonumber\\
	&+S \epsilon_{+} -\tfrac12 \gamma_{0}\dot{F}_3\e_-\ , 
 \nonumber\\
	\delta S= & -\tfrac12 \bar{\epsilon}_{+} \gamma_0 \dot{\psi}_1 +\tfrac12 \bar{\epsilon}_{+} \gamma^i \partial_i \psi_2  - \tfrac12 \bar{\e}_-\gamma_{0}\dot{\psi}_2\ .
\end{align}

\vspace{0.5cm}
	\section{Tensor Calculus for the Construction of a $\cN=2$ Carroll Supersymmetric Scalar Field Theory }
	
	Based on the supersymmetry transformation rules, is possible to write the components of the primed scalar multiplet in terms of $(\phi, \lambda, F)$ as
	\begin{eqnarray}
		\phi^\prime &=& F \ddot \phi + \tfrac12 \dot{\bar{\l}} \dot \l \,,\nonumber\\
		\l^\prime &=&   F \ddot \l+ \dot F \dot \l  +  \tfrac12 \ddot \phi \g^i \partial_i \l - \tfrac12 \g^i \partial_i \dot \f \dot \l \,,\nonumber\\
		F^\prime &=&  F \ddot F + \dot F^2 + \tfrac14 \ddot \f \partial^i \partial_i \f - \tfrac14 \partial_i \dot \f \partial^i \dot \f \nonumber\\
		&& - \tfrac12 \ddot{\bar \l} \g^i \partial_i \l  - \tfrac12 \dot{\bar \l} \g^i \partial_i \dot \l \,.
  \label{CompositeScalar}
	\end{eqnarray}
Here the lowest component was chosen in this particular way to achieve the $Q_-$-invariance, and the remaining composite expressions are obtained by employing the supersymmetry transformation rules on both sides of the equation. Then, using the composite scalar multiplet \eqref{CompositeScalar}, along with the invariant action formulae
	\begin{eqnarray}
		\cL &=& \phi F^\prime + \phi^\prime F - \bar\l \l^\prime \,.
	\end{eqnarray}
	the $\cN=2$ Carroll supersymmetric scalar field theory with spatial derivatives is given by
	\begin{eqnarray}
		\cL &=& \f F \ddot F + \f\dot F^2 + \tfrac14 \f \ddot \f \partial^i \partial_i \f - \tfrac14 \f\partial_i \dot \f \partial^i \dot \f +  F^2 \ddot \phi\nonumber\\
		&& - \tfrac12 \f \ddot{\bar \l} \g^i \partial_i \l - \tfrac12 \f \dot{\bar \l} \g^i \partial_i \dot \l  +  \tfrac12 F \dot{\bar{\l}} \dot \l  -  F  \bar\l \ddot \l \nonumber\\
		&& - \dot F  \bar\l \dot \l  -  \tfrac12 \ddot \phi  \bar\l \g^i \partial_i \l + \tfrac12 \bar\l \g^i \partial_i \dot \f \dot \l  \,.
	\end{eqnarray}
	Performing a sequence of partial integrations, along with a rescaling of the Lagrangian with an overall factor of $-4/3$, the Lagrangian simplifies as follows
	\begin{equation}
		\cL = -2 F^2 \ddot \phi +\tfrac12 \dot\f^2 \partial_i \partial^i \f -2 F \dot{\bar \l} \dot \l -2 \dot \f \dot{\bar\l} \g^i \partial_i \l \,.
  \end{equation}
This action represents the supersymmetric completion of the Carroll Galileon model $\cL \sim  \phi ( \ddot\f\,\partial^i \partial_i \f - \partial_i \dot \f \partial^i \dot \f )$.

\end{document}